\documentclass[letterpaper, 9pt]{article}
\usepackage{geometry}
\usepackage[numbers,round,sort&compress,comma]{natbib}

\usepackage{times,mathptmx,amsfonts, amsmath, bm, bbm, graphicx, epsfig, epstopdf}
\usepackage{siunitx}
\usepackage{esvect}

\usepackage{sectsty}
\usepackage{balance}
\usepackage{lipsum,lineno} 
\usepackage{color}

\usepackage{lastpage}

\usepackage{fancyhdr}
\usepackage{graphicx}
\usepackage{etoolbox}

\usepackage{tocloft}  
\usepackage[nottoc,numbib]{tocbibind} 

\cftpagenumbersoff{section}
\cftpagenumbersoff{subsection}
\cftpagenumbersoff{subsubsection}

\usepackage{indentfirst}

   \makeatletter
   \def\@seccntformat#1{\@ifundefined{#1@cntformat}%
   {\csname the#1\endcsname\quad}
   {\csname #1@cntformat\endcsname}
   }
   \def\section@cntformat{\thesection.\quad}
   \def\subsection@cntformat{\thesubsection.\quad}
   \def\subsubsection@cntformat{\thesubsubsection.\quad}
   \def\thebibliography#1{\section{References\@mkboth
   		{REFERENCES}{REFERENCES}}\list
   	{[\arabic{enumi}]}{\settowidth\labelwidth{[#1]}\leftmargin\labelwidth
   		\advance\leftmargin\labelsep
   		\usecounter{enumi}}
   	\def\newblock{\hskip .11em plus .33em minus .07em}
   	\sloppy\clubpenalty4000\widowpenalty4000
   	\sfcode`\.=1000\relax}
   \makeatother

\newcommand{\remove}[1]{}
\begin{document}

\makeatletter

\def\paragraph{\@startsection{paragraph}{4}{10pt}{-1.25ex plus -1ex minus -.1ex}{0ex plus 0ex}{\normalsize\textit}}
\renewcommand\@makefntext[1]%
{\noindent\makebox[0pt][r]{\@thefnmark\,}#1}
\makeatother
\renewcommand{\figurename}{\small{Fig.}~}
\sectionfont{\large}
\subsectionfont{\normalsize}

\noindent\LARGE{\textbf{Are there optical communication channels in the brain?}}
\vspace{0.6cm}

\noindent\large{\textbf{Parisa Zarkeshian\textit{$^{1}$}, Sourabh Kumar\textit{$^{1}$}, Jack Tuszy{\'n}ski\textit{$^{2,3}$}, Paul Barclay\textit{$^{1,4}$}, and Christoph Simon \textit{$^{1}$}}}\vspace{0.5cm}

\noindent\small\textit{$^{1}$~Institute for Quantum Science and Technology and Department of Physics
	and Astronomy, University of Calgary, Calgary T2N 1N4, Alberta, Canada}\\
\noindent\small\textit{$^{2}$~Department of Oncology, University of Alberta, Cross Cancer Institute, Edmonton T6G 1Z2, Alberta, Canada}\\
\noindent\small\textit{$^{3}$~Department of Physics, University of Alberta, Edmonton  T6G 2E1, Alberta, Canada}\\
\noindent\small\textit{$^{4}$~National Institute for Nanotechnology, Edmonton T6G 2M9, Alberta, Canada}

\renewcommand\contentsname{\textbf{TABLE OF CONTENTS}}
\noindent\small\textit{\tableofcontents} 

\vspace{0.5cm}

\section{ABSTRACT}
Despite great progress in neuroscience, there are still fundamental unanswered questions about the brain, including the origin of subjective experience and consciousness. Some answers might rely on new physical mechanisms. Given that biophotons have been discovered in the brain, it is interesting to explore if neurons use photonic communication in addition to the well-studied electro-chemical signals. Such photonic communication in the brain would require waveguides. Here we review recent work [S. Kumar, K. Boone, J. Tuszynski, P. Barclay, and C. Simon, Scientific Reports 6, 36508 (2016)] suggesting that myelinated axons could serve as photonic waveguides. The light transmission in the myelinated axon was modeled, taking into account its realistic imperfections, and experiments were proposed both \textit{in vivo} and \textit{in vitro} to test this hypothesis. Potential implications for quantum biology are discussed.
	
\section{INTRODUCTION}

Over the past decades a substantial number of facts has been discovered in the field of brain research. However, the fundamental question of how neurons, or more specifically all particles involved in the biological processes in the brain, contribute to mental abilities such as consciousness is still unanswered. The true explanation to this question might rely on physical processes other than those that have been discovered so far. One interesting candidate to focus on is biophotons, which might serve as supplementary information carriers in the brain in addition to the well established electro-chemical signals.

Biophotons -- which are photons ranging from near-IR to near-UV frequency and emitted without any enhancement or excitation-- have been observed in many organisms such as bacteria \cite{bacteria}, fungi \cite{fungi}, germinating seeds \cite{GSs}, plants \cite{plants}, animal tissue cultures \cite{ATC},  and different parts of the human body \cite{HB1,HB2,HB3,HB4}, including the brain \cite{isojima,kobayashi,tang1, kataoka,salari2016,daiPNAS}. These biophotons are produced by the decay of electronically excited species which are created chemically during oxidative metabolic processes \cite{chang2013biophotons, cifra2014ultra} and can contribute to communication between cells \cite{fels2009cellular}. Moreover, several experimental studies show the effects of light on neurons' and, generally, the brain's function \cite{wade,starck,leszkiewicz}. The existence of biophotons and their possible effects on the the brain along with the fact that photons are convenient carriers of information raises the question whether there could be optical communication in the brain.

For the sources and detectors of the optical communication process in the brain, mitochondrial respiration \cite{zhuravlev,tuszynski1} or lipid oxidation \cite{mazhul}, and centrosomes \cite{buehler} or chromophores in the mitochondria \cite{kato} have been proposed, respectively. It has also been observed that opsins, photoreceptor protein molecules, exist in the brains of birds \cite{okano,nakane}, mammals \cite{blackshaw,nissila,kojima2011,provencio}, and more general vertebrates \cite{kojima} and even in other parts of their bodies \cite{tarttelin,shichida} as well.

Another essential element for this optical communication, which is not well established yet, is the existence of physical links to connect all of these spatially separated agents in a selective way. In the dense and (seemingly) disordered environment of the brain, waveguide channels for traveling photons would be the only viable way to achieve the targeted optical communication processes. Mitochondria and microtubules in neurons have been introduced as the candidates for such waveguides \cite{thar,rahnama,jibu1,scholkmann}. However, they are not suitable in reality due to their small and inhomogeneous structure  for light guidance over proper distances in the brain.

\begin{figure*}[t]
	\includegraphics[scale=0.7]{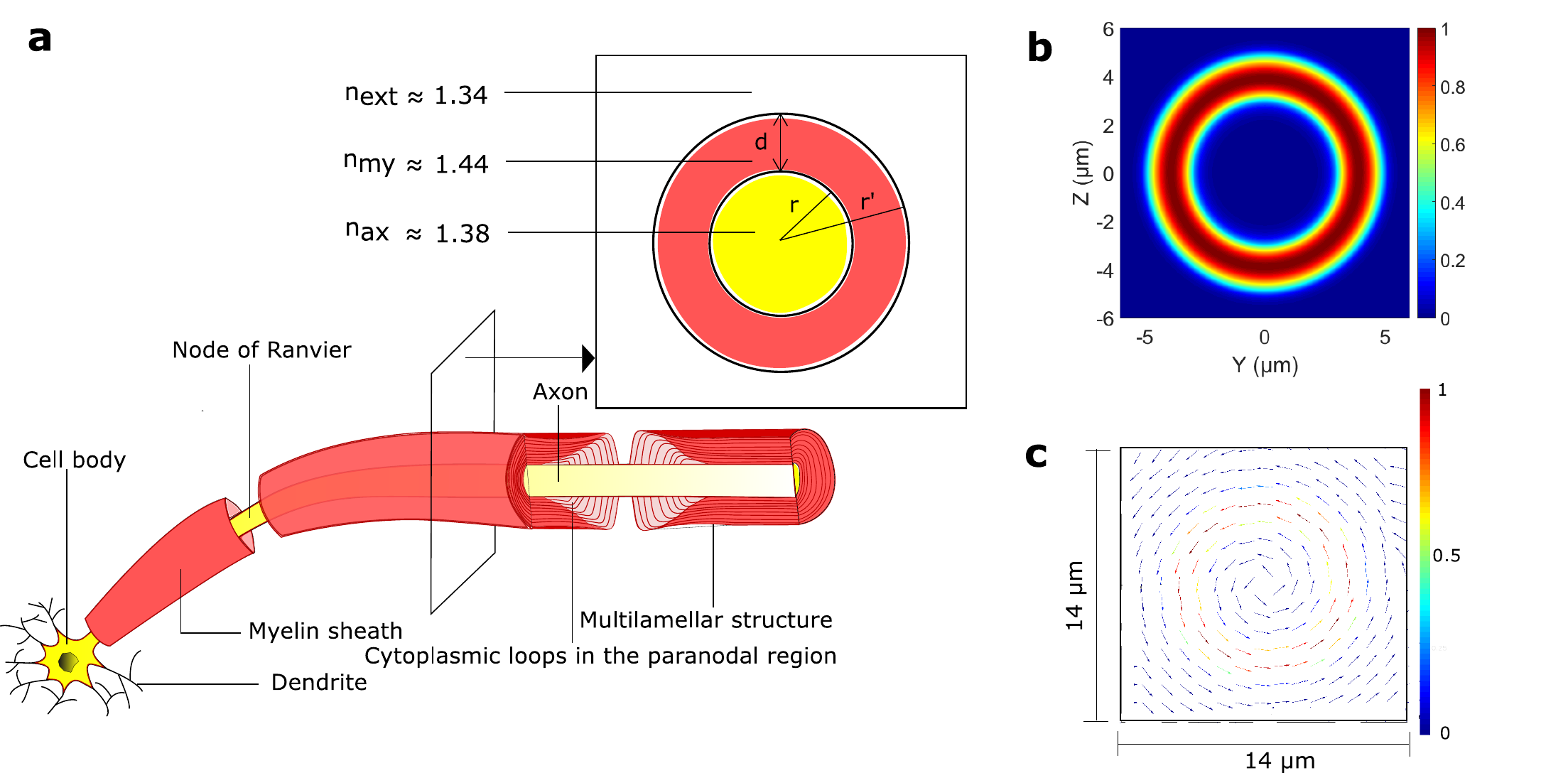}
	\caption{\textbf{Simplified depiction of a segment of a neuron, and the cylindrically symmetric eigenmode of a myelinated axon.}
	\textbf{(a)} Structure of a segment of a neuron which is cut longitudinally near the end of the segment. Each layer of the compact myelin sheath (shown in red) ends in the cytoplasm filled loops (shown in light red) in the paranodal region close to a Node of Ranvier. The inset indicates the cross section in the transverse plane with \textit{r} and \textit{r}$'$ as the inner and outer radii of the myelin sheath. \textit{d} is the thickness of the myelin sheath, and $\mathrm{n_{my}}$, $\mathrm{n_{ax}}$, and $\mathrm{n_{ext}}$ are the refractive indices of the myelin sheath, the inside of the axon, and the interstitial fluid outside, respectively. \textbf{(b)} Electric field magnitude of a cylindrically symmetric eigenmode (with wavelength $\mathrm{\lambda}$ = \SI{0.612}{\micro\meter}) for a myelinated axon (with \textit{r}=\SI{3}{\micro\meter}, and \textit{r}$'$=\SI{5}{\micro\meter}). \textbf{(c)} Electric field vector at different points for displaying the azimuthal polarization of the input mode. The adjacent color bar shows the field magnitude. Figure from Ref. \cite{kumar}.} 
	\label{schematic_and_input}
\end{figure*}

Ref. \cite{kumar} proposed myelinated axons as potential biophoton waveguides in the brain. The proposal is supported by a theoretical model and numerical results taking into account real imperfections. Myelin sheath (formed in the central nervous system by a kind of glia cell called oligodendrocyte) is a lamellar structure surrounding the axon and has a higher refractive index \cite{antonov} than both the inside of the axon and the interstitial fluid outside (see Fig.~\ref{schematic_and_input}a) which let the myelin sheath to guide the light inside itself for optical communications.  This compact sheath also increases the propagation speed of an action potential (via saltatory conduction) based on its insulating property \cite{purves}. There has been a few indirect experimental evidence for light conduction by axons \cite{tang1,sun,hebeda}. Another related and interesting experiment has shown that a certain type of glia cells, known as M\"{u}ller cells, guide light in mammalian eyes \cite{franze, labin}. Ref. \cite{kumar} also proposed experiments to test the existence of the optical waveguides in the brain.

One interesting property of optical communication channels is that they can also transmit quantum information. Quantum effects in biological systems are being studied in different areas such as photosynthesis \cite{engel,romero}, avian magnetoreception \cite{ritz,hiscock}, and olfaction \cite{turin,franco}. There is an increasing number of conjectures about the role of quintessential quantum features such as superposition and entanglement \cite{horodecki} in the brain \cite{daiPNAS,jibu1,mashour,hameroff,fisher}. The greatest challenge when considering quantum effects in the brain or any biological system in general is environmentally induced decoherence \cite{tegmark1}, which leads to the suppression of these quantum phenomena. However, some biological processes can be fast and may
show quantum features before they are destroyed by the environment. Moreover, nuclear spins can have coherence times of tens of milliseconds in the brain \cite{warren,lei}. A recent proposal on ``quantum cognition'' suggests even longer coherence times of nuclear spins \cite{fisher}, but relies on quantum information transmission via molecule transport, which is very slow. In contrast, photons are the fastest and most robust carriers for quantum information over long distances, which is why currently man-made quantum networks rely on optical communication channels (typically optical fibers) between spins \cite{kimble,sangouard}.

\section{Results}

To show that myelinated axons could serve as the waveguides for traveling biophotons in the brain, Ref. \cite{kumar} solved the three dimensional electromagnetic field equations numerically in different conditions, using Lumerical's software packages FDTD (Finite Difference Time Domain) Solutions and MODE Solutions. These software packages solve Maxwell's equations numerically, allowing the optical properties of dielectric structures defined over a mesh with subwavelength resolution to be simulated.

The refractive indices of the fluid outside of the axon, the axon, and the myelin sheath were taken close to 1.34, 1.38 and 1.44 respectively (see Fig.~\ref{schematic_and_input}a), which are consistent with their typical values \cite{antonov, wang, tuchin}. These indexes let the myelin sheath guide the light inside itself. The ratio of the radius of the axon, \textit{r} to the outer radius of the myelin sheath \textit{r}$'$ (\textit{g-ratio}) is taken equal to 0.6 for the most of the simulations, close to the experimental values \cite{friede}. In reality, the radius of the myelinated axons in the brain changes from $0.2$ microns to close to $10$ microns \cite{liewald}. For the purpose of guiding light inside the myelin sheath, Ref. \cite{kumar}
considered the wavelength of the observed biophotons in the brain which is from $200$ nm to $1300$ nm. Since several proteins in the environment of the axons  strongly absorb at wavelengths close to $300$nm, a wavelength range of the transmitted light from the shortest permissible  wavelength, $\lambda_{min}=400$nm, to the longest one, $\lambda_{max}$, was chosen to avoid the absorption and confine the light well in the myelin sheath. $\lambda_{max}$ is chosen to the upper bound of the observed biophoton wavelength (\SI{1300}nm) or the thickness of the myelin sheath (denoted by $d$), whichever is smaller. Besides $\lambda_{min}$ and $\lambda_{max}$,  an intermediate wavelength was considered, denoted by $\lambda_{int}$, corresponding to the central permissible frequency (mid-frequency of the permissible frequency range) in the simulations.

In the following section we discuss the guided modes in the myelinated axons and their transmissions in nodal and paranodal regions and even in the presence of the imperfections such as bends, varying cross-sections, and non-circular cross-sections.\\

\subsection{Optical transmission in myelinated axons}\hfill

Within the neuron, one can identify numerous intra-cellular structures that can function as potential scatterers, i.e.\ sources of waveguide loss. They are located both inside the axon and outside of the axon. Intra-cellular structures include cell organelles, for example, mitochondria, the endoplasmic reticulum, lipid vesicles, as well as the many filaments of the cytoskeleton, namely microtubules, microfilaments and neurofilaments. Extra-cellular structures include microglia, and astrocytes. However, the electromagnetic modes which are spatially confined within the myelin sheath, should not be affected by the presence of these structures. These biophoton modes considered here would be able to propagate in a biological waveguide provided its dimension is close to or larger than the wavelength of the light.
Fig.~\ref{schematic_and_input}b shows the numerically calculated magnitude of the electric field of a cylindrically symmetric eigenmode of an axon with radius $r=3\mu$m and myelin sheath radius $r'=5\mu$m for the wavelength $0.612\mu$m. This electric field is azimuthally polarized as depicted is Fig.~\ref{schematic_and_input}c and it is similar to the $\mathrm{TE_{01}}$ mode of a conventional fiber \cite{jocher} which has higher refractive index of the core than that of the cladding. It is important to note that azimuthal polarization would prevent modal dispersion in the birefringent myelin sheath.  Importantly, its optical axes are oriented radially \cite{chinn}. It can be readily established that there are hundreds of potential guided modes allowed to exist given the thickness of myelin sheath. Consequently, biophotons that could be generated by a source in the axons (e.g. mitochondria or recombination of reactive oxygen species) could readily interact with these modes as determined by mode-specific coupling coefficients. While we lack detailed knowledge of the particulars for these interactions, for the sake of simplicity and ease of illustration we select a single mode and examine its transmission. It is interesting to analyze transmission in the presence of optical imperfections such as discontinuities, bends and varying cross-sectional diameters.  In this connection, we simulated short axonal segments due to computational limitations and extrapolated the results for the full length of an axon.\\

\begin{figure}[ht]
	\centering
	\includegraphics[scale=0.27]{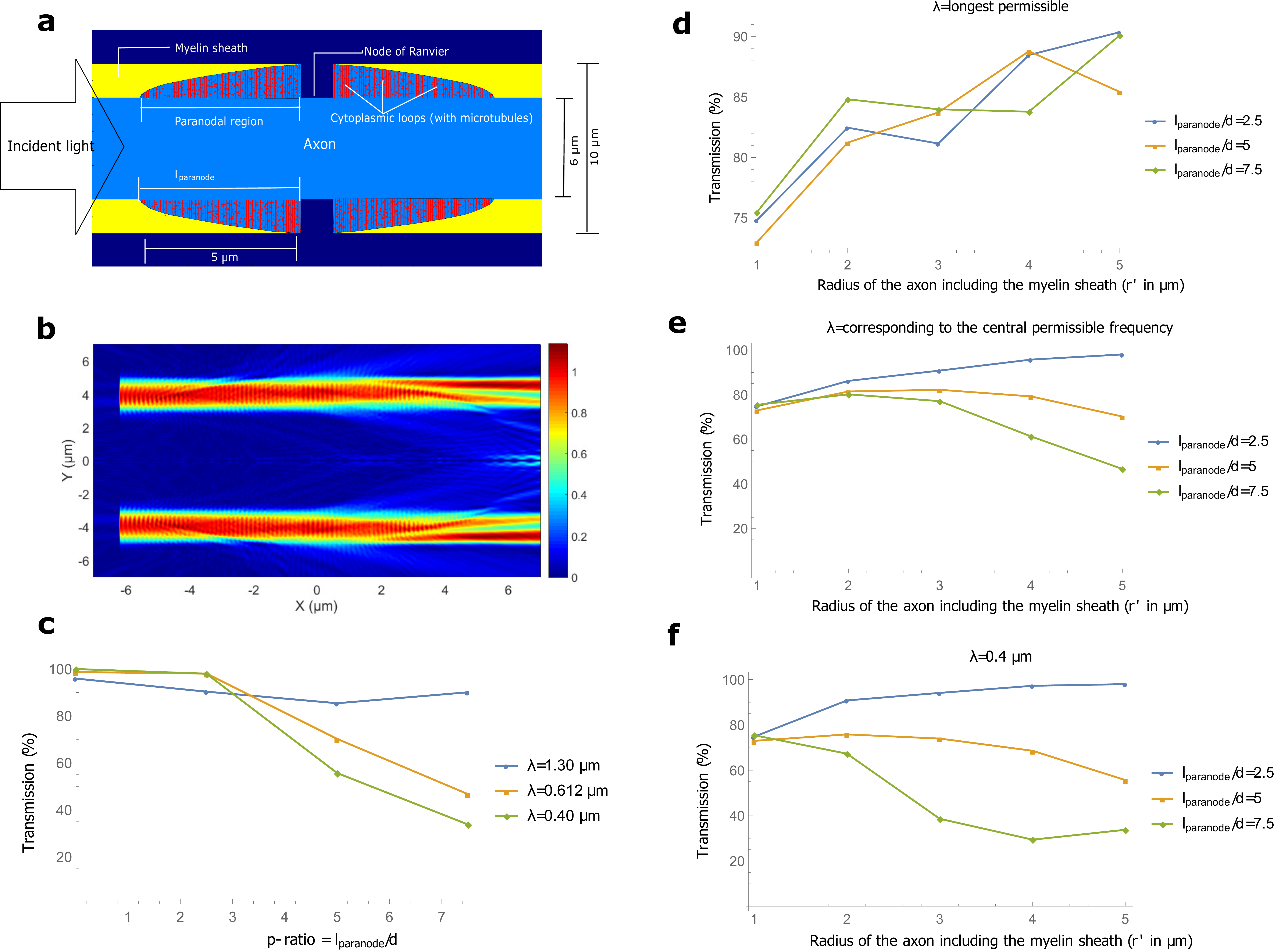}
	\caption{\:\:\textbf{Nodal and paranodal regions and their transmissions.} \textbf{(a)} Longitudinal cross-section of the nodal and  paranodal regions in the model of Ref. \cite{kumar}. Here, the ratio of the radius of the axon and the outer radius of the myelin sheath, \textit{g-ratio}, is equal to 0.6, where the outer radius of the myelin sheath is chosen $r'$ = \SI{5}{\micro\meter}, and the length of the paranode, $l_{paranode}$ = \SI{5}{\micro\meter}. \textbf{(b)} Magnitude of the electric field profile in the longitudinal direction (EFPL) when a cylindrically symmetric input mode with wavelength \SI{0.612}{\micro\meter} passes through the nodal and paranodal region. \textbf{(c)} Transmission percentage of an axon with the outer radius $r'$ = \SI{5}{\micro\meter}, as a function of the \textit{p-ratio} (the ratio of the paranodal length and the thickness of the myelin sheath). \textbf{(d)}-\textbf{(f)} Transmission percentage as a function of the axon radius for three different wavelengths of the input mode and three different lengths of paranode. Figure from Ref. \cite{kumar}.}
	\label{paranode}
\end{figure}
\subsubsection{Transmission in nodal and paranodal regions}\hfill

A myelinated axon has periodically unmyelinated segments, called Nodes of Ranvier, which are approximatly $1\mu$m long \cite{kandel} (while the whole axon length varies from $1$ mm to the order of a meter). Here, we discuss the transmission in the Ranvier nodes and at the edges of the nodes, the paranodes. The configuration of myelin sheath is special in the paranodal regions (see Fig.~\ref{paranode}a). There are many layers making up the compact myelin sheath and at the edge of each node, almost all of the layers are in contact with the core (bared axon) with a small pocket of cytoplasm. That's because each layer moving from the innermost outward is longer than the one below. However, for thick myelin sheaths, many cytoplasmic pockets cannot reach the surface of the bare axon, but end on inner layers. Thus, the length of paranodal regions is dependent on the thickness of the myelin sheath. We call the ratio of the length of paranode, $l_{paranode}$, to the thickness of the myelin sheat, $d$, \textit{p-ratio} and take its value close to 5 in our simulations based on the realistic values \cite{zagoren}.

Fig.~\ref{paranode}a displays the model of Ref. \cite{kumar} for two adjacent paranodal regions with the node in between, and Fig.~\ref{paranode}b shows the magnitude of the electric field profile in the longitudinal direction (along the length of the axon), EFPL, as a cylindrically symmetric input mode crosses this region.
Fig.~\ref{paranode}c shows the power transmission in the guided modes as a function of \textit{p-ratio} for three wavelengths, \SI{0.40}{\micro\meter}, \SI{0.61}{\micro\meter}, and \SI{1.30}{\micro\meter}. For the transmission, there are two main losses: divergence or scattering of the light beam. Shorter wavelengths scatter more but diverge less.
Thus, in Fig.~\ref{paranode}c, for small \textit{p-ratio}s, shorter wavelengths have higher transmission and as the effect of divergence is dominant in this region and the shorter wavelengths diverge less. However, for the large \textit{p-ratio}s, the effect of scattering is dominant and since the higher wavelengths scatter less, and have a higher transmission.

Fig.~\ref{paranode}d--f compares the transmission percentage for different axon radii, wavelengths, and \textit{p--ratios}. Although in Fig.~\ref{paranode}d, the  behavior of the transmission as a fuction of axon radius is independent of \textit{p--ratios} for the longest permissible wavelength, it can be concluded that for the most loosely confined modes ($\lambda_{max}$) transmission increases in thicker axons. It's also possible that for long wavelengths, a fraction of the light diverging into the axon comes back into the myelin sheath at the end of the paranodal region and not all the light that diverges is lost. This can be an explanation for not well-defined dependency of the transmission on the paranodal lengths (see Fig.~\ref{paranode}d).

In, Fig.~\ref{paranode}e, and Fig.~\ref{paranode}f, for \textit{p-ratio} = 2.5, based on our intuition from Fig.~\ref{paranode}c, the divergence is dominant. Here, the thickness of the axon plays a role in the transmission such that the thicker the axon the divergence is less and the light is transmitted more. However, for larger \textit{p--ratios}, the scattering is dominant and the light scatters more in thick axons.

To summarize, for small \textit{p--ratios} ($\sim$2.5), the well confined modes (shorter wavelengths) transmit better while for large \textit{p--ratios} ($\sim$5 or greater), the loosely confined ones (longer wavelengths) transmit better. Thicker axons yield higher transmission for all wavelengths with small \textit{p--ratios} while it's inverse only for the shorter wavelengths with large \textit{p--ratios}. The transmission after several paranodal regions can be roughly estimated by following the intuition of exponentiating the transmission through one (see Supplementary Information of Ref \cite{kumar}).\\

\subsubsection{Transmission in bends}\hfill

\begin{figure}[ht]
	\begin{center}
		\includegraphics[scale=0.23]{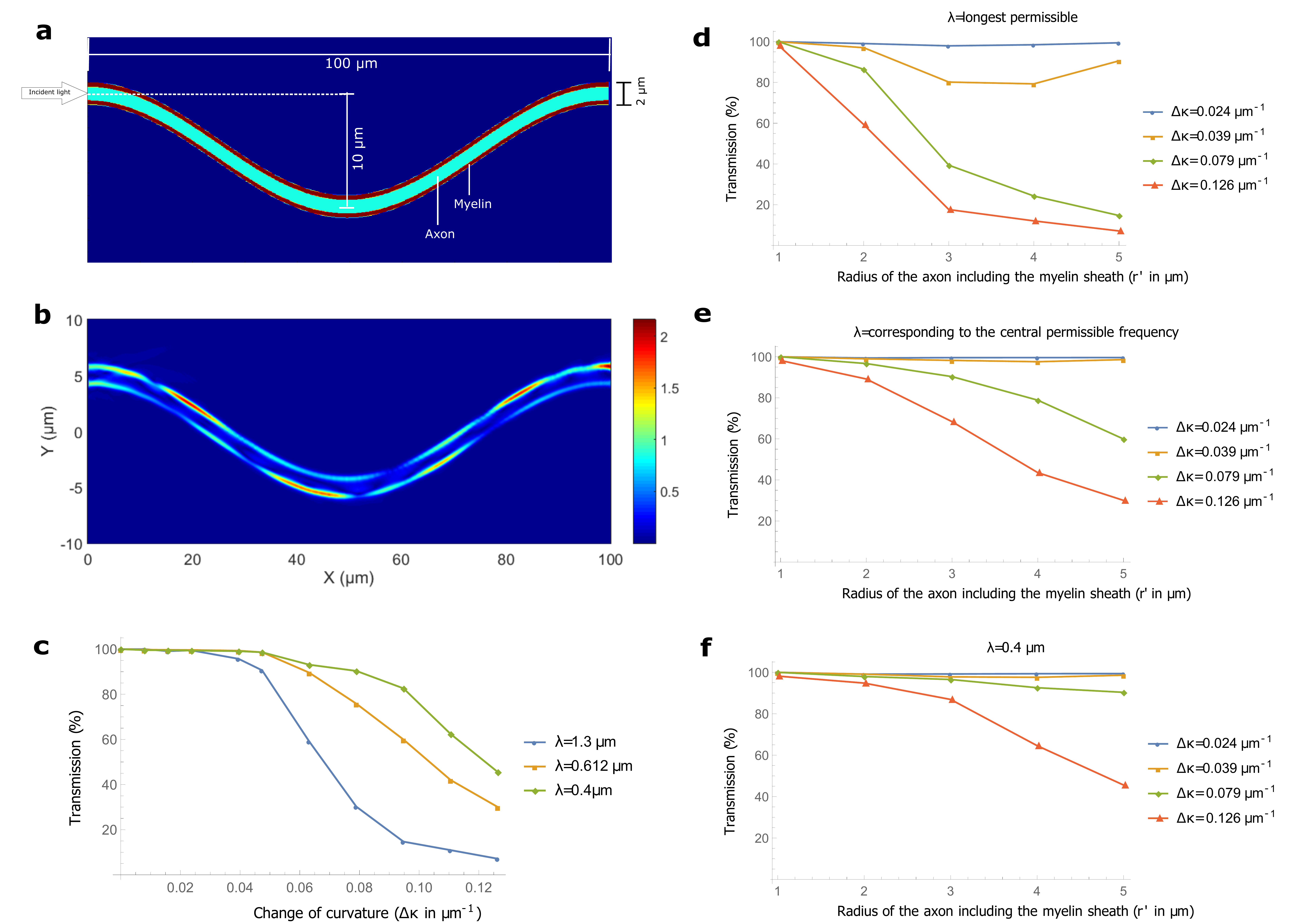}
		\caption{\:\:\textbf{Bends in the waveguides and their transmission.} \textbf{(a)} a schematic of a sinusoidally bent waveguide in our model. Here, $r'$ = \SI{1}{\micro\meter}, the wavelength of the cosine function is \SI{100}{\micro\meter}, and its amplitude (\textit{A}) is \SI{5}{\micro\meter}. \textbf{(b)} Magnitude of the EFPL when the input mode with wavelength \SI{0.4}{\micro\meter} passes through the bent waveguide. \textbf{(c)} Transmission percentage as a function of the change of curvature, $\Delta \kappa$, for three different wavelengths where $r'$ =  \SI{5}{\micro\meter} and $A$ is being changed to vary $\Delta \kappa$. \textbf{(d)}-\textbf{(f)} Transmission percentage as a function of the myelinated axon radius for three different wavelengths with four different $\Delta \kappa$. Figure from Ref. \cite{kumar}.}
		\label{bend}
	\end{center}
\end{figure}

Power transmission of a straight waveguide has loss on encountering bends in the waveguide. Although this type of loss can be minimized by propagation of the eigenmodes of circular bends of constant curvature along the waveguide, one cannot use them for axons. For the varying curvature of axons, these modes are more lossy than eigenmodes of a straight waveguide. So, to verify the bend losses in the axons, Ref. \cite{kumar} considered the straight--mode in a sinusoidal waveguide with changing curvature and obtained the transmission in the myelin sheath at the other end.
Fig.~\ref{bend}a shows one example of S-bend in an axon with radius \SI{0.6}{\micro\meter}, and Fig.~\ref{bend}b shows the EFPL as a straight--mode passes through the axon.

The loss in sinusoidal S-shaped bends (shown in Fig.~\ref{bend}a) is highly dependent on the varying curvature \cite{syahriar}. Thus, Ref. \cite{kumar} calculated the total power transmission up to a wavelength away from the myelin sheath boundaries as a function of the change of curvature, $\Delta \kappa = 4 A k^2$  (\textit{k} is the wavenumber and $A$ is the amplitude of the cosine function) and plotted it for 3 different wavelengths in Fig.~\ref{bend}c ($r'$=\SI{5}{\micro\meter}). The shorter wavelength gives higher transmission and the more the curvature changes the smaller the transmission. Fig.~\ref{bend}d--f compares the transmission of the 3 different wavelengths for different radii of axon and different curvatures. For $\Delta \kappa$ $\sim$\SI{0.024}{\micro\meter^{-1}}, all the permissible wavelengths are guided with transmission percentage close to $100\%$ for all axon radii. Note that here we consider the change of curvature, $\Delta\kappa$, of the curve passing through the central axis of the axon. However, the inner part of the bent has the most curvature in comparison with the center or outer part at each point. These differences are more noticeable for thicker axons since they have more change of curvature than the thiner ones and therefore experience more loss for the same $\Delta\kappa$. For the typical axons in the brain similar to those in the images of Ref \cite{schain} (the relatively straight axons with length of $\sim$\SI{1}{\milli\meter} and radius of $\sim$\SI{1}{\micro\meter}), $\Delta \kappa<$ \SI{0.05}{\micro\meter^{-1}} which results in transmission of over 90 $\%$.\\

\begin{figure}[ht]
	\begin{center}
		\includegraphics[scale=0.21]{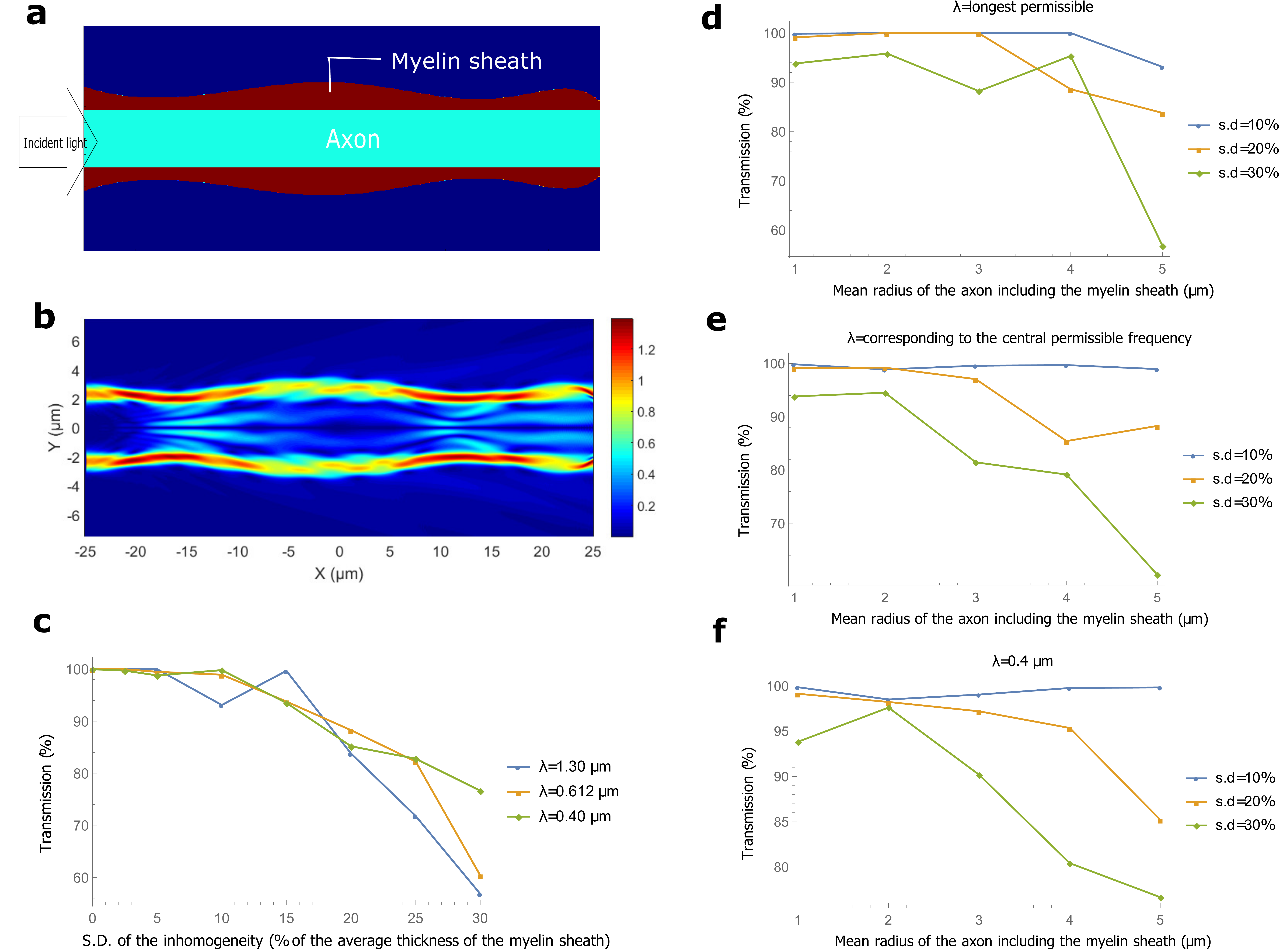}
		\caption{\:\:\textbf{Varying cross-sections and the transmission in such areas.} \textbf{(a)} a schematic of a myelinated axon with longitudinally varying cross sections. Here, the mean radius of the myelinated axon (outer radius of the axon) is \SI{3}{\micro\meter} and the standard deviation (s.d.) of the variation of radius for the myelin sheath is \SI{0.36}{\micro\meter}. \textbf{(b)} Magnitude of the EFPL when the input mode with wavelength \SI{0.48}{\micro\meter} passes through the region. \textbf{(c)} Transmission percentage as a function of the s.d.\ of the variation in radius for the myelin sheath for three different wavelengths (the mean radius of the myelinated axon is chosen as \SI{5}{\micro\meter}). \textbf{(d)}-\textbf{(f)} Transmission percentage as a function of the mean radius of the myelinated axon for three different wavelengths with three different s.d.\ of the variation of  radius for the myelin sheath. Figure from Ref. \cite{kumar}.}\label{ca}
	\end{center}
\end{figure}

\subsubsection{Transmission in presence of varying cross-sections}\hfill

The thickness of the myelin sheath, \textit{d}, is not uniform all along the length of the axon. Ref. \cite{kumar} varied \textit{d} according to an approximate normal distribution. The correlation length in the roughness of the myelin sheath -- the width of bumps or valleys in the outer surface of myelin sheath-- was taken to be \SI{5}{\micro\meter} to \SI{10}{\micro\meter}. The mean value of the distribution ($r'$) is chosen based on the value of \textit{g-ratio}, and the standard deviation (s.d.) of \textit{d} is varied.
Fig.~\ref{ca}a shows an example of the simulation for an axon with length of \SI{50}{\micro\meter}, \textit{r}=\SI{2.4}{\micro\meter}, and the s.d.\ of 30  $\%$ of the average \textit{d}. In Fig.~\ref{ca}b, the EFPL for input light with $\lambda = $ \SI{0.612}{\micro\meter} is calculated. Fig.~\ref{ca} c shows that in general, more deviation of thickness of the myelin sheath results in less transmission and shorter wavelengths transmit better than longer ones. In Fig.~\ref{ca}d--f, we see the behavior of the transmission as a function of mean radius of the myelinated axon with 3 different deviation and for 3 different wavelengths. For variations less than 10$\%$, almost all of the wavelengths can pass through with efficiency over 90$\%$.  These results were obtained for the \SI{50}{\micro\meter} segment of the axon and was extrapolated to the case of a longer segment transmission along the axon. In particular, the transmission fraction was exponentiated by the number of \SI{50}{\micro\meter} segments formed in the axon. In general, thicker axons are more sensitive to the large deviations and suffer more loss. Note that longer correlation lengths lead to better transmission for the same s.d, while significantly shorter correlation lengths are known to strongly scatter the mode. Some of the axonal segments (length $\sim$\SI{5}{\micro\meter}) of thin axons (\textit{r} $\sim$\SI{1}{\micro\meter}) are within this type of inhomogeneity, as represented in the images of \cite{peters}. However, we were unable to find appropriate images of thicker myelinated axons and longer segments required for a more realistic estimate of this type of inhomogeneity.\\

\subsubsection{Transmission in non-circular cross sections}\hfill

\begin{figure}[ht]
	\begin{center}
		\includegraphics[scale=0.27]{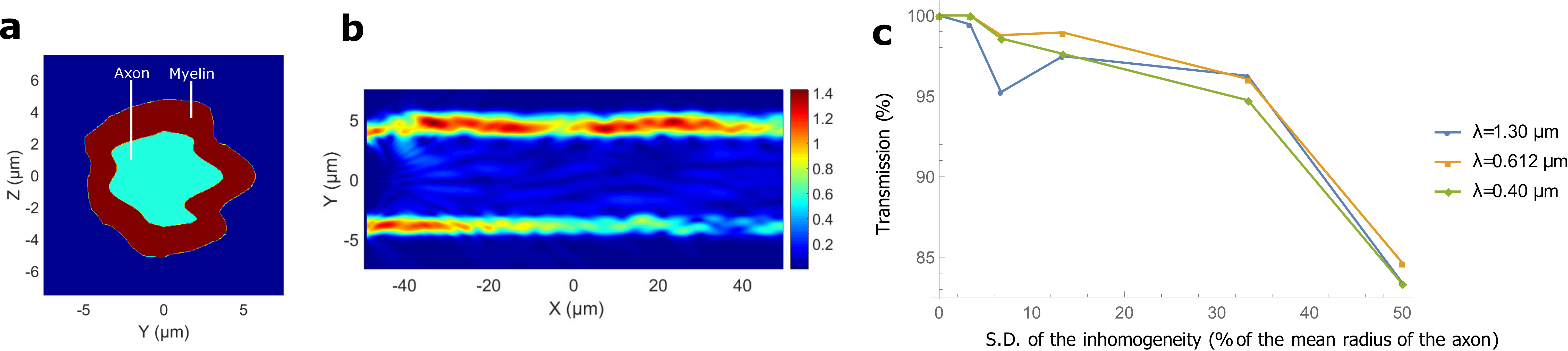}
		\caption{\:\:\textbf{Non-circular cross sections of the axon and myelin sheath and their transmission.} \textbf{(a)} A model of the cross-section of an axon with the myelin sheath. The mean distance of the points along the cross-sectional boundary of the axon from its center is \SI{3}{\micro\meter} with s.d.\ of \SI{0.4}{\micro\meter}.
		A parallel curve with an approximate of \SI{2}{\micro\meter} apart from the axon's boundary is taken as the myelin sheath's boundary.
	    \textbf{(b)} Magnitude of the EFPL when a cylindrically symmetric input mode for a circular cross-section passes along the waveguide with the modeled non-circular cross-section. Here, the input mode wavelength is \SI{0.612}{\micro\meter}, the axon radius is \SI{3}{\micro\meter}, and the myelinated axon's radius is \SI{5}{\micro\meter}. \textbf{(c)} Transmission percentage as a function of the s.d.\ of the distance between the points on the boundary of the axon and a circle of radius \SI{3}{\micro\meter} for three different wavelength. Figure from Ref. \cite{kumar}.}
		\label{nccs}
	\end{center}
\end{figure}

Axons and their myelin sheaths can have different cross-sectional shapes \cite{peters}. As an example, Ref. \cite{kumar} simulated the cross-section
of a myelinated axon shown in Fig.~\ref{nccs}. The points along the axon’s cross-sectional boundary in Fig.~\ref{nccs}a follow a normal distribution whose mean value is $3 \mu$m and standard deviation $0.4 \mu$m. The myelin sheath is approximated as a parallel curve drawn at a perpendicular distance of $2 \mu$m giving an average \textit{g-ratio} $= 0.6$. Fig.~\ref{nccs}b displays the magnitude of the EFPL for the incident light of $\lambda=612 nm$. As Fig.~\ref{nccs}c shows, the total power transmission decreases for all wavelengths while the cross-section becomes more random. But the effect of this loss is small in reality, as many axons have less than $10\%$ inhomogeneity in the cross-sectional shape. Therefore, if the axon and myelin sheath change their cross-sectional shape slightly along the length of the axon, the primary source of loss would be the coupling loss \cite{peters}. However, for a substantial change of circular cross-section along the length, there will be propagation loss as well \cite{kumar}.\\

\subsubsection{Transmission in presence of other imperfections}\hfill

In addition to the sources of loss considered so far, there can be more imperfections. One of the notable ones is the cross-talk between axons. The conducted light can leak from one axon to the other one if they are placed so close to each other. To avoid this loss, the distance between two adjacent axon should be at least a wavelength, which happens in most of the realistic cases based on the images in Ref.~\cite{peters}.

The other imperfection we discuss is the varying refractive indexes of the axon and the myelin sheath which have been assumed constant until now. Changing the refractive index both transversely and longitudinally with a s.d.\ of $0.02$ (typical variation as expected from \cite{antonov, wang}), while keeping the mean the same as the one used so far, Ref. \cite{kumar} observed no considerable difference in the transmission (typically less than 1 $\%$). Furthermore, Ref. \cite{kumar} considered the effect of the glia cells next to the internodal segment. These cells were modeled as spheres with radii varying from \SI{0.1}{\micro\meter} to \SI{0.3}{\micro\meter}, and refractive index $1.4$, filling up one third of the volume of the nodal region outside the axon as expected from the images of the Ref.~\cite{peters}. For the thickest axons, transmission through a node of Ranvier was not affected significantly while for the thinnest one, the transmission increased slightly ($\sim$2 $\%$). There would be also additional scattering losses in the areas that myelin sheath is inhomogeneous. \\

\subsection{Absorption}

In biological tissues, and more so in the brain, scattering of light, rather than absorption, is the main source of attenuation of optical signals \cite{cheong}. To our knowledge, the absorption coefficient of the myelin sheath has not been measured experimentally. One can only infer it indirectly with limited accuracy. The average absorption coefficient in the white matter decreases almost monotonically from $\sim$\SI{0.3}{\milli\meter^{-1}} to $\sim$\SI{0.07}{\milli\meter^{-1}} for wavelengths \SI{0.4}{\micro\meter} to \SI{1.1}{\micro\meter} \cite{yaroslavsky}. But myelin can not be responsible for the majority of the absorption since grey matter (which is almost devoid of myelin) has comparable absorption coefficients \cite{yaroslavsky}. It is likely that light sensitive structures (e.g.\ chromophores in the mitochondria) are the main contributors to the absorption. Another way to infer myelin's absorption coefficient is to look at the absorption of its constituents, i.e.\ lipids, proteins and water. Mammalian fat shows an absorption coefficient less than \SI{0.01}{\milli\meter^{-1}} for the biophotonic wavelength range \cite{veen}. Water has similar absorption coefficients. Most proteins have a strong resonance peak close to \SI{0.28}{\micro\meter} with almost negligible absorption above \SI{0.34}{\micro\meter}, and the proteins in the myelin (e.g. myelin protolipid protein, and myelin basic protein) behave similarly \cite{facci}. Thus, absorption in myelin for the biophotonic wavelengths seems negligible (over a length scale of $\sim$\SI{1}{\centi\meter}), based on the data of its constituents.

\subsection{Attainable transmission}

Ref. \cite{kumar} estimated the attainable transmission percentage after $1$cm length of the axon in different examples, taking the axon diameter as 100--150 times less than its internodal length according to the realistic values \cite{friede,ibrahim}. For an axon with $r$ = \SI{3}{\micro\meter}, $r'$ = \SI{5}{\micro\meter}, \textit{p-ratio} = 7.5, internodal length = \SI{1}{\milli\meter}, wavelength of input light = \SI{1.3}{\micro\meter}, s.d.\ for varying area = 2.5 \%, $\Delta \kappa $ = \SI{0.039}{\micro\meter^{-1}}, s.d.\ for non-circularity in cross-section shape = 13.33 \%, and separation from the nearby axons = \SI{1}{\micro\meter}, the transmission after 1 cm would be $\sim$31 \%. This transmission can be increased to $\sim$82 \% if we take the wavelength of input light = \SI{0.61}{\micro\meter}, $\textit{p-ratio}$ = 2.5, and keep all the other parameters the same. For a thinner axon with $r$ = \SI{1.8}{\micro\meter}, $r'$ = \SI{3}{\micro\meter}, $\textit{p-ratio}$ = 7.5, internodal length = \SI{500}{\micro\meter}, wavelength of light = \SI{1.2}{\micro\meter}, s.d.\ for varying area = 20 \%, separation from other axons = \SI{1.2}{\micro\meter}, and $\Delta \kappa $ = \SI{0.039}{\micro\meter^{-1}} would yield $\sim$3 \% transmission after 1 cm.

If one chooses the shorter length of \SI{2}{\milli\meter} for the axons (as there are axons with $\sim$\SI{1}{\milli\meter} length in the brain\cite{nolte}), then the transmission for the 3 examples discussed above would be $\sim$78 \%, $\sim$96 \%, and $\sim$46 \%, respectively. The main source of loss for these examples is the coupling in the paranodal regions. By locating the sources and receivers close to the ends of the myelinated sections of the axon, one can reduce coupling losses. It is worth noting that photons can propagate in either directions: from the axon terminal up to the axon hillock or in the opposite direction along the axon.\\

\subsection{Attainable communication rates}
To estimate the biophoton emission rate per neuron, one can use the experimental data of Ref.~\cite{tang1} in which the number of biophotons emitted per minute by a slice of mouse brain is counted while the neurons are excited with the neurotransmitter glutamate. Ref. \cite{kumar} calculated this biophoton emission rate as about 1 photon per neuron per minute. This rate is about one to two orders of magnitude lower than the the average rate of electrochemical spikes \cite{buzsaki}. But there is a considerable uncertainty on this number and it might be higher in reality since it only takes into account biophotons scattered outside, while most of them would likely be absorbed in the brain itself rather than being scattered (if they were propagating inside the waveguides in the brain). On the other hand, one can argue that the estimate could also be too high because the brain slice was stimulated with glutamate. One should also notice that this rate can be different depending on the neuron type.

If one takes such low rate of biophoton emission and consider the fact that there are about $10^{11}$ neurons in a human brain, there would still be over a billion photon emission per second. This mechanism appears to be sufficient to facilitate transmission of a large number of bits of information, or even allow the creation of a large amount of quantum entanglement. Note that the behavior of about one hundred photons can already not be simulated efficiently with classical computers \cite{lloyd}. It is also worth to mention that psychophysical experiments performed in the past indicate that the bandwidth of conscious experience is below the range of 100 bits per second \cite{zimmermann,norretranders}. \\

\subsection{Proposals to test the hypothesis}

Although there is already some experimental evidence of biophoton propagation in the brain and axons \cite{tang1,sun,hebeda}, it would nevertheless be very interesting to test the light guidance of axons directly \textit{in-vitro} and \textit{in-vivo}.
For testing \textit{in-vitro}, one way is to light up one end of a thin brain slice (with proper homogeneous myelinated axons) and look for the bright spots related to the open ends of the myelinated axons at the other end. To get more accurate results, one can isolate a suitable myelinated axon and, while keeping it alive in the proper solution, couple into one of guided modes of the axon, similar to what was done for light guidance by M\"{u}ller cells in Ref. \cite{franze}. Since the real light sources in the brain may be close to axons terminals \cite{tang1}, one can cut the axon near the terminal and hillock regions, couple the mode directly to the myelin sheath and observe the light intensity from the other end of the axon. This test can verify the guidance of the myelin sheath. Evanescent coupling and readout of light offers another possibility.

For an \textit{in-vivo} test of light guidance, first one needs to prove the existence of biophotons in the myelin sheath. To do so, one can add light-sensitive chemicals like $\mathrm{AgNO_3}$ into the cytoplasmic loops  in the paranodal region or in oligodendrocytes, which will be passed around to the myelin sheath. Then, the light-activated decomposition of $\mathrm{AgNO_3}$ to $\mathrm{Ag}$ leaves $\mathrm{Ag}$ atoms as the dark insoluble grains. The idea of this test is similar to the technique involved in the development of photographic films, and the in-situ biophoton autography (IBA) technique \cite{sun}.

An additional type of tests that can be done \textit{in-vivo} is to insert fluorescent molecules or nano-particles as the sources. They could also serve as the detectors if their fluorescence emission is due to the absorption of photons emitted from the sources \cite{yang}. One can also employ optogenetics \cite{deisseroth} to create artificial detectors. Optogenetics uses neurons which are genetically modified to produce light-sensitive proteins operating as ion-channels (like channel rhodopsin). If one places these light-sensitive proteins into the axonal membrane at the end of myelin sheath close to an axon terminal or into the the membranes of the cytoplasmic loops in the paranodal region, one can detect light by observing the operation of the ion-channels. It is worth noting that the rate of oligodendrogenesis increases and myelin sheath becomes thicker in neighboring of light-illuminated genetically modified neurons \cite{gibson}. The question comes up whether the neurons adjust themselves for a better light guidance by forming more layers.

Besides the artificial sources and detectors, it is also interesting to use natural ones to show the photon guidance from the sources to the detectors through the axons. To do so, one needs to first understand well the photon sources, and their emission rates and wavelength. One can make use of nanoantennas to raise the emission rates \cite{kuhn} and get a better knowledge about the sources and emissions. Although, some measurements on the number of scattered photons from the axons (not guided or absorbed ones) \cite{isojima, kobayashi, tang1, kataoka} have been performed, photon emission from the individual neurons has not yet been analyzed. One also needs to measure photon detection capabilities of the natural candidate detectors such as opsins, centrosomes \cite{buehler} and chromophores in mitochondria \cite{kato}. \\

\section{Discussion}

In this review of Ref. \cite{kumar} we have discussed how light conduction in a myelinated axon is feasible even in the presence of realistic imperfections in the neuron. We have also described future experiments that could validate or falsify this model of biophoton transmission \cite{kumar}. It is also worth addressing a few related questions. It is of interest to identify possible interaction mechanisms between biophotons and nuclear spins within the framework of quantum communication.  Spin chemistry research \cite{spin-chemistry} determined effects whereby electron and  nuclear spins affect chemical reactions. These effects can also involve photons. In particular, a class of cryptochrome proteins can be photo-activated resulting in the production of a pair of radicals per event, with correlated electronic spins.  This effect has been hypothesized to explain bird magnetoreception \cite{ritz}. It has been recently shown by theoretical considerations that interactions between electron and nuclear spins in cryptochromes are of critical importance to the elucidation of the precision of  magnetoreception effects \cite{hiscock}. Importantly for this topic, cryptochrome complexes are found in the eyes of mammals and they are also magnetosensitive at the molecular level \cite{foley}. Therefore, if similar proteins can be found in the inner regions of the human brain, this could provide the required interface between biophotons and nuclear spins. However, for individual quantum communication links to form a larger quantum network with an associated entanglement process involving many distant spins, the nuclear spins interfacing with different axons must interact coherently. This, most likely, requires close enough contact between the interacting spins. The involvement of synaptic junctions between individual axons may provide such a proximity mechanism.

We should also address the question of the potential relevance of optical communication between neurons with respect to consciousness and the binding problem. A specific anatomical question that arises is whether brain regions implicated in consciousness \cite{koch-review} (e.g. claustrum \cite{koubeissi, crick}, the thalamus, hypothalamus and amygdala \cite{loewenstein}, or the posterior cerebral cortex \cite{koch-review}) have myelinated axons with sufficient diameter to allow light transmission.

A major role of the myelin sheath as an optical waveguide could provide a better understanding of the causes of the various diseases associated with it (e.g. multiple sclerosis \cite{nakahara}) and hence lead to a design and implementation of novel therapies for these pathologies.

Let us note that, following Ref. \cite{kumar}, we have focused our discussion here on guidance by myelinated axons. However, light guidance by unmyelinated axons is also a possibility, as discussed in more detail in the supplementary information of Ref. \cite{kumar}.

Finally, with the advantages optical communication provides in terms of precision and speed, it is indeed a wonder why biological evolution would not fully exploit this modality.  On the other hand, if optical communication involving axons is harnessed by the brain, this would reveal a remarkable, hitherto unknown new aspect of the brain’s functioning, with potential impacts on unraveling fundamental issues of neuroscience.\\

\section{Acknowledgment}

P.Z. is supported by an AITF scholarship. S.K. is supported by AITF and Eyes High scholarships. J.T., P.B. and C.S. are supported by NSERC.

\vspace{1mm}

\noindent\textbf{Key Words:} Brain, Biophotons, Myelinated axons, Waveguides, Quantum, Review\\

\noindent\textbf{Send correspondence to:} Christoph Simon, Department of Physics and Astronomy, University of Calgary, Calgary T2N 1N4, Alberta, Canada, Tel: +1(403)220-7007, Fax: +1(403)210-8876, E-mail: csimo@ucalgary.ca\\

\end{document}